\documentclass[preprint, 12pt]{amsart}
\usepackage[table]{xcolor}
\usepackage{times}
\usepackage{amssymb, amsmath}
\usepackage{amsthm}
\usepackage{tikz}
\usetikzlibrary{matrix,arrows}
\usepackage{enumitem}
\usepackage{booktabs}
\usepackage{color}
\usepackage{multirow}

\theoremstyle{plain}

\theoremstyle{remark}

\definecolor{Gray}{gray}{0.95}
\newcolumntype{g}{>{\columncolor{Gray}}c}
\graphicspath{{./images/}}

\begin{document}
\title[Stock price forecasting]{Forecasting significant stock price changes using neural networks}

\author{Firuz Kamalov}

\address{ Canadian University Dubai, Dubai, UAE.}
\email{\textcolor[rgb]{0.00,0.00,0.84}{firuz@cud.ac.ae}}

%\date{\today}
\date{\today
\newline \indent $^{*}$ Corresponding author}

\begin{abstract}
Stock price prediction is a rich research topic that has attracted interest from various areas of science. The recent success of machine learning in speech and image recognition has prompted researchers to apply these methods to asset price prediction.
The majority of literature has been devoted to predicting either the actual asset price or the direction of price movement. In this paper, we study a hitherto little explored question of predicting \textit{significant} changes in stock price based on previous changes using machine learning algorithms. We are particularly interested in the performance of neural network classifiers in the given context. To this end, we construct and test three neural network models including multi-layer perceptron, convolutional net, and long short term memory net. As benchmark models we use random forest and relative strength index methods. The models are tested using 10-year daily stock price data of four major US public companies. 
Test results show that predicting significant changes in stock price can be accomplished with a high degree of accuracy. In particular, we obtain substantially better results than similar studies that forecast the direction of price change.
\end{abstract}

\maketitle

%-----------------------------------------------------------------------------------------------------------------------------------------------------
%-----------------------------------------------------------------------------------------------------------------------------------------------------
\section{Introduction}
The ability to predict future stock prices has been the holy grail of many inside the financial industry as well as academia. The implications of being able to correctly forecast stock prices have fueled interest in the topic from the early days of stock markets. However, the work of Fama \cite{fama1} and subsequently Fama-French \cite{fama2} put a damper on these efforts. Fama argued convincingly that stock prices contain all publicly available information implying that stock price prediction is a fruitless endeavor. Despite these discouraging reports researchers and analysts have continued trying to develop stock price prediction models using various approaches. In the quest to find novel solutions researchers drew on behavioral science, physics, genetics, and other fields for inspiration \cite{agustini, karathan, nguyen}.  The recent success of machine learning in speech and image recognition has encouraged researchers to turn their attention to artificial intelligence \cite{guresen, heaton}. The majority of the attempts to predict asset prices have focused on the actual price or price direction. Our goal is to employ modern machine learning models to forecast \textit{significant} changes in asset price. Concretely, we use daily returns over previous $p$ days to predict if the next daily return would be significant.

Economic data is large and complex so it is extremely difficult to delineate its complicated inner relationships with an econometric model. Machine learning models are universal algorithms that are able to capture complex nonlinear relationships within data which makes them appealing for use in financial modeling. A great deal of effort has been directed to applying machine learning to stock price prediction though with varying degrees of success.  One of the most convincing examples of success of quantitative analysis and machine learning in finance is the amazing performance of the Medallion Fund over the past two decades  \cite{gergaud}. Nonetheless, machine learning algorithms must be applied with caution as most remain black box models.

Stock price prediction is often done in one of the two ways: numerical price prediction or price direction prediction. In numerical price prediction, a learning model such as a regression is built to predict the actual price of a stock. In direction prediction, a learning model such as a classifier is built to predict the direction - up or down - of  price movement. The former task remains a daunting challenge with most of the modern quantitative methods unable to beat a simple random walk model in out of sample testing \cite{fama2}. The latter task appears more feasible \cite{nelson} as it requires less precision - the direction of price change contains less information than the actual price change. Note that regression models can also be adapted to predict the direction of price change by considering only the sign of the predicted output. Although information about the direction of price change does not tell the whole story it is still immensely useful and profitable. 

In this paper, we extend the idea of predicting the direction of price change to predicting significant changes in price. While small changes in the direction of price can happen frequently, significant changes in price are more rare and are driven by different fundamentals. 
Clearly some of the significant movements in price are due to unexpected news that would be impossible to forecast. On the other hand, it seems plausible that we can learn to identify situations where a stock is oversold or overbought which would lead to a reversal in price change.
In our forecast models, we employ sophisticated deep learning algorithms such as convolutional neural networks (CNN) and long term short memory (LSTM) networks. We contrast the performance of CNNs and LSTMs with multilayer perceptron (MLP) which is a plain feedforward neural network. We use random forest (RF) to add a non-neural net machine learning algorithm to our study. RF is  a simple and efficient classification algorithm that serves as a great benchmark model.

Price change indicators have existed in finance long before the advent of machine learning. Relative strength index (RSI) is one such popular financial statistic that is used to identify oversold or overbought stocks. RSI is calculated based on the closing prices over a recent trading period. According to Wilder, RSI index above 70 or below 30 indicates an overbought or oversold stock respectively \cite{wilder}. RSI has stood the test of time and remains in use in both industry and academia  \cite{gurrib, sahin}.
We use RSI in our study to compare the performance of machine learning models to the traditional finance methods. 

We use the daily stock price data of four major US companies over a 10-year period to build the forecast models. We analyze the performance of the models in predicting significant positive and negative daily returns. The results indicate that machine learning models are more successful at forecasting significant daily returns achieving AUC of almost 0.85 in some cases (Figure \ref{csco}). In addition, all the tested learning models substantially outperformed the RSI model.   

The paper is divided as follows. In Section 2, we briefly review the existing literature on stock market prediction using machine learning. In Section 3, we describe the algorithms used in the study. In Section 4, we present our experiments and results. We end the paper with concluding remarks in Section 5. 

\section{Literature}

Machine learning has recently experienced great success in areas such as image and speech recognition \cite{szegedy, xiong}. As a result researchers became encouraged to apply the same techniques to build financial forecasting models \cite{heaton, nelson}. 
The authors in \cite{fischer} carried out a large scale study applying LSTM on the constituents of S\&P 500 index between 1992-2015. Results showed that LSTM based approach outperforms other machine learning approaches in predicting out-of-sample directional movements.
In  \cite{chen}, the authors applied LSTM to predict returns in Chinese stock market. The results showed a 13\% improvement in accuracy over random prediction method.
The authors in \cite{bao} proposed a novel approach to forecasting next day stock prices by using a three stage procedure. In the first stage the time series data is denoised using wavelet transform followed by feature extraction using auto-encoders and applying LSTM in the final stage. The proposed model produced better results than other similar models in accuracy and profitability performance.
An ensemble of LSTMs was used in \cite{borovkova} to predict the intraday change in direction of stock prices for 22 large cap US stocks. The authors engineered a set of basic and advanced input features for the networks to enhance the performance of the models. The weighted ensemble model performed consistently, albeit marginally, better than benchmark lasso and ridge logistic models.

Ensemble techniques that combine several machine learning methods have been actively explored in the literature.  In \cite{chatzis}, a suite of  learning methods was used to model the probability of stock market crash event during various time frames. The authors showed that  deep neural networks significantly increase the classification accuracy. The authors in \cite{liew} tested random forests, support vector machines, and deep neural networks to predict returns on ETFs. Concretely, the authors used prior returns, trading volume, and dummy variables to predict the direction of price change of ETFs. The results showed that the methods work best over three to six-month horizons. In addition, trading volume is found to be a strong predictor of future return direction. The authors in \cite{patel} used a combination of machine learning methods to predict stock market indexes in India. In the first stage, the authors applied  support vector regression to predict values of technical parameters on day $t+n$ based on input values from day $t$. The output from Stage 1 was then used as input for Stage 2 models that included support vector regression, artificial neural networks, and random forest. Experiments showed that the two-stage models have better accuracy than single-stage models in predicting stock index. 

Combining traditional econometric models with modern machine learning tools has become another popular approach albeit with mixed results. The authors in \cite{kim} combined various GARCH type models with LSTM to forecast stock price volatility. Experimental results on Korean stock exchange index data revealed that the hybrid  GEW-LSTM model outperformed standalone models such as GARCH, ANN, and LSTM. In \cite{sun2}, the authors use ARMA-GARCH together with artificial neural networks to create an intelligent system for predicting stock market shocks. The results suggest that the proposed model can effectively predict market shocks based on intraday trading data. On the other hand, the study by  \cite{guresen} on stock index prediction shows that the basic multi-layer perceptron outperforms more involved neural networks. The authors compared the performance of multi-layer perceptron with dynamic and hybrid neural networks using daily values of NASDAQ composite index. The results indicate that the more complex neural network architectures do not necessarily lead to better performance.

%----------------------------------------------------------------------------------------------------------------------------------------------------
%----------------------------------------------------------------------------------------------------------------------------------------------------

\section{Machine learning models}
Neural networks are a class of machine learning algorithms that are patterned after the neurons inside human brain. There exist many variations of neural networks starting with MLP and ending with more exotic architectures such as LSTM. Neural networks achieved their recent success due to three main factors: novel and improved architectures, increase in computational power stemming from the use of GPUs, and creation of large training datasets. The early neural networks such as MLP, although powerful, did not quite outperform other machine learning methods such as support vector machines and random forests. One of the the first major breakthroughs took place with introduction of CNNs which were used by Lecun \cite{lecun} to achieve high accuracy in classification of handwritten digits. Subsequent deep learning models such as Alexnet \cite{krizhevsky} and Resnet \cite{he} that consist of tens and hundreds of hidden layers and trained on millions of samples pushed image classification accuracy to even greater levels. The success of neural networks thrust their application in a wide array of fields beyond image recognition.

The main distinguishing characteristic of neural networks is their ability to `learn' new features from data. In case of image recognition, neural networks can learn to identify edges, shapes, and outlines which are then combined to label the image. In applying neural networks to stock prices we hope that they learn hidden features or patterns in the data that would lead to correct price prediction. In our study, we employ three of the most popular types of neural networks: MLP, CNN, and LSTM. Each network has its own flavor thus providing us with a broad overview of this class of machine learning algorithms.

\subsection{Multilayer Perceptron}
\begin{figure}[h!]
\center
\includegraphics[scale=0.5]{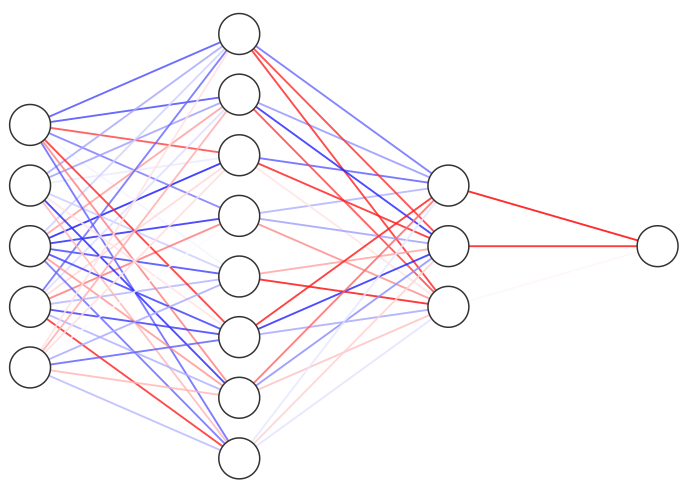}
\caption{Multi-layer perceptron architecture.}
\label{mlp}
\end{figure}

Multilayer perceptron (MLP) is a  basic type of feedforward neural network that consist of an input layer, hidden layer(s), and an output layer (Figure \ref{mlp}). Each layer consists of a number of nodes which are interconnected via weights. During the model training stage the algorithm adjusts the weights of the network to increase classification accuracy. The model training consists of several forward and backward passes. In the forward pass, the data is passed through the network from the input layer to the output layer. In the backward pass, the algorithms calculates partial derivatives of the cost function with respect to the weights and uses them to adjust the values of the weights. Despite its relatively basic structure MLP remains an effective model for classification \cite{dudek}.

During the training phase a single forward pass consists of calculating node values of successive layers starting with the input layer. 
The number of nodes in the input layer corresponds to the number of input features. The input features are fed into the nodes of the first hidden layer. The weighted sum of the input values plus a bias term is then transformed using a nonlinear function such as \textit{sigmoid},\textit{ tanh} or \textit{ReLu}.  This process is continued until the output layer node(s) is calculated. Thus, in a certain sense, a MLP is nothing more than a composition of a series of affine transformations and certain nonlinearities. 
The cost function in a classification task is defined based on mutual information between the predicated and actual values of the target variable. 

In a backward pass of the training stage the network weights are adjusted according to the corresponding partial derivatives of the cost function. This corresponds to a single gradient descent step in minimizing the cost function. The partial derivatives are calculated in reverse order starting from the output layer. The chain rule is used to calculate partial derivatives for the weights of successive layers  based on previous layers. The use of chain rule greatly simplifies derivative calculations which makes MLP an appealing algorithm.

\subsection{Convolutional Neural Network}
\begin{figure}[h!]
\center
\includegraphics[scale=0.5]{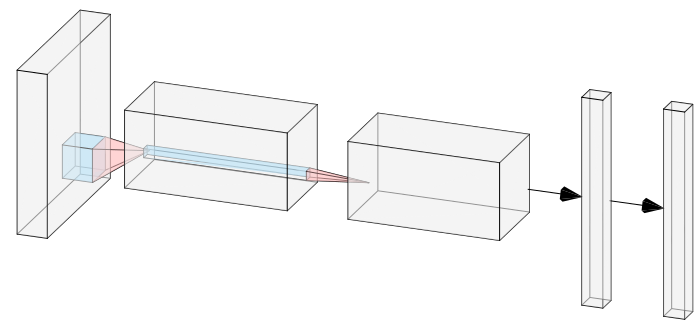}
\caption{Convolutional neural network architecture in the style of AlexNet.}
\label{cnn}
\end{figure}

Convolution is a popular mathematical tool that is used in computer science and engineering. The idea for convolutional neural networks (CNN) was motivated by the use of convolution in image processing. CNN architecture in many ways resembles that of MLP. The main distinguishing characteristic of CNN are convolutional layers. A convolutional layer is calculated by sliding a window (filter) across an input array and taking the dot product with the corresponding part of the array (Figure \ref{cnn}). In this way, CNN takes advantage of any existing structure within the input data. CNNs often contain pooling layers that are used to refine the signal that is passed through the network. Thus, a CNN usually consists of several convolution and pooling layers followed by traditional dense layers. There exist several popular CNN architectures such AlexNet, ResNet, and Inception whose success in image recognition has made deep learning a state-of-the-art machine learning method. Since time series data is inherently structured CNN is a good candidate to exploit any underlying patterns.

\subsection{Long Short Term Memory}
\begin{figure}[h!]
\center
\includegraphics[scale=0.5]{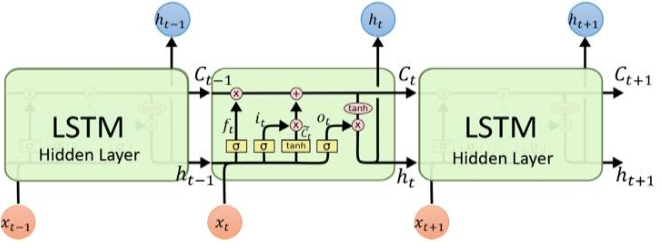}
\caption{LSTM architecture (Source: http://colah.github.io/posts/2015-08-Understanding-LSTMs/)}
\label{lstm}
\end{figure}

LSTM is a type of a recurrent neural network that has been used successfully in natural language processing. Recurrent neural networks (RNN) were designed to process sequential data - consisting of multiple time steps. In a typical RNN the output is calculated based on the current  input and the  previous hidden state, where the hidden state is calculated during the previous time step. Thus the network `remembers' previous inputs as it calculates the current output. A regular RNN suffers from the vanishing gradient phenomenon whereby the gradient value rapidly decreases as it propagates back in time. A small gradient means that the weights of the initial layers will not be updated effectively during the training session. LSTM solves the vanishing gradient problem by introducing an LSTM unit into a regular RNN. An LSTM unit consists of three gates - input gate, forget gate, and output gate - that control the flow of information inside the unit (Figure \ref{lstm}). 
LSTMs have been shown to perform well on sequential data. Therefore, they are inherently well suited for time-series analysis. 

\subsection{Random Forest}
Random forest is a classical machine learning tool that is based on aggregating the output of a collection of decision trees \cite{breiman}. Thus RF reduces overfitting that is characteristic of individual decision trees. Each decision tree is constructed by recursively splitting data at different values of input features. The choice of the split is determined based on the corresponding information gain. The main advantage of a decision tree is speed and interpretability. However, decision trees tend to overfit the data. To reduce overfitting a bootstrap aggregation technique is applied. The data is repeatedly sampled and the corresponding decision tree is constructed. Then the output of the bootstrap model is determined by taking the mode of outputs of individual trees. In RF individual decision tree have an additional property that at each split only a subset of all features is considered. It is done to reduce the correlation among the trees and thereby reduce the output variance.

\subsection{Relative Strength Index}
RSI is a popular financial indicator used to gauge the degree to which an asset is being oversold or overbought in the market \cite{wilder}. It is calculated based the ratio of average gains to average losses over a trailing 14-day period. An RSI value of under 30 indicates that the stock is oversold. Similarly, an RSI over 70 indicates that the stock is overbought. We use this simple logic as the predictive model for significant changes. Concretely, we predict a significant negative change when RSI reaches 70 or above. And we predict a significant positive change when RSI reaches 30 or under.

%----------------------------------------------------------------------------------------------------------------------------------------------------
%----------------------------------------------------------------------------------------------------------------------------------------------------

\section{Numerical Experiments}
In this section we present the results of our experiments that were carried out to test the performance of machine learning algorithms in predicting significant changes in stock price. The results indicate that machine learning tools - and in particular neural networks - can be used effectively to forecast significant price changes in stock price. 

\subsection{Methodology}
In our experiments, we test three major neural network models: MLP, CNN, and LSTM. The models are built using the TensorFlow library. In order to maintain comparability we use the same general architecture in all three models (Table \ref{nn_arch}). Each model consists of an input layer, two hidden layers, and an output layer. We use ReLu activation function in every layer except the output layer. Dropout rate of 0.2 is applied to certain layers in CNN and LSTM models. In addition to neural networks we also use a RF model to represent more traditional learning algorithms. The RF model is imported from the scikit-learn library with its default settings. To benchmark the performance of machine learning algorithms we use a RSI based predictive model. It is a widely used financial indicator that signals when an asset is potentially oversold (overbought). RSI is calculated using Wilder's moving average with different size lookback windows.  

\begin{table}[h!]
\centering
\caption{Neural network architectures}
\label{nn_arch}
\begin{tabular}{@{}lll@{}}
\toprule
\textbf{Network} & \textbf{Hidden Layer 1}    & \textbf{Hidden Layer 2}    \\ \midrule

MLP     & Dense 64     & Dense 32     \\ \hline
CNN     & \begin{tabular}[c]{@{}l@{}}Conv1D 64,\\ length 7,\\ dropout rate = 0.2\end{tabular}          & \begin{tabular}[c]{@{}l@{}}Dense 32,\\ dropout rate = 0.2\end{tabular} \\ \hline

LSTM    & \begin{tabular}[c]{@{}l@{}}LSTM 64 with\\ return sequence,\\ dropout rate = 0.2\end{tabular} & \begin{tabular}[c]{@{}l@{}}LSTM 32,\\ dropout rate = 0.2\end{tabular}  \\ \bottomrule
\end{tabular}
\end{table}

The experiments are performed using data on four major US publicly traded companies: Coca-Cola, Cisco Systems, Nike, and Goldman Sacks. We use adjusted daily stock prices from 2009 to 2019. The data is converted to daily returns  prior to the experiments. The daily return is calculated based on the following formula:

$$r_t = \ln\big(\frac{p_t}{p_{t-1}}\big),$$
where $r_t$ and $p_t$ indicate the return and price for day $t$ respectively. To ensure the integrity of the experiments the data is split temporally into training and testing parts using a 75\%/25\% ratio. Input feature vectors consist of prior returns over the previous $p$ days and output is the current return value, i.e. 
$$\boldsymbol{x}_k = (r_{k-p}, r_{k-p+1}, ..., r_{k-1}), \,\, y_k = r_k.$$ 
The experiments are performed using a range of values for $p$: 7, 14, 30, and 60 days. A daily return value is defined as significant if it exceeds a predefined threshold. The threshold is calculated as a fraction of the standard deviation of daily returns over the training period. Thus, when the fraction is 1.2 any return value over the threshold of 1.2$\sigma$ is considered as positively significant, where $\sigma$ is the standard deviation of daily returns in the training set. We carry out experiments using a range of fraction values to observe the effects of varying threshold levels on the performance of classifiers. 

Since significant daily returns constitute a small portion of all returns our data has an imbalanced class distribution which can affect the performance of classifiers. Class imbalance can result in a biased classifier whereby the majority class points are given  preference over minority samples. A common approach to addressing this issue is through resampling the minority data to achieve a balanced distribution. We leave addressing this problem to future research.

As mentioned above class imbalance is an important issue in the context of our study. In particular, the choice of classifier performance metric requires consideration. Since classifier's goal is to increase accuracy it will often do so at the expense of minority instances. Therefore, using accuracy or error rate would not reflect the true performance of a classifier. Area under ROC curve (AUC) is often used to remedy this issue. The ROC curve is obtained by plotting the true positive rate of a classifier against the false positive rate at different threshold levels. Thus, AUC represents the probability that a classifier will rank a randomly chosen positive instance higher than a randomly chosen negative instance.

\subsection{Results}
\begin{figure}[h!]
\center
\includegraphics[scale=0.42]{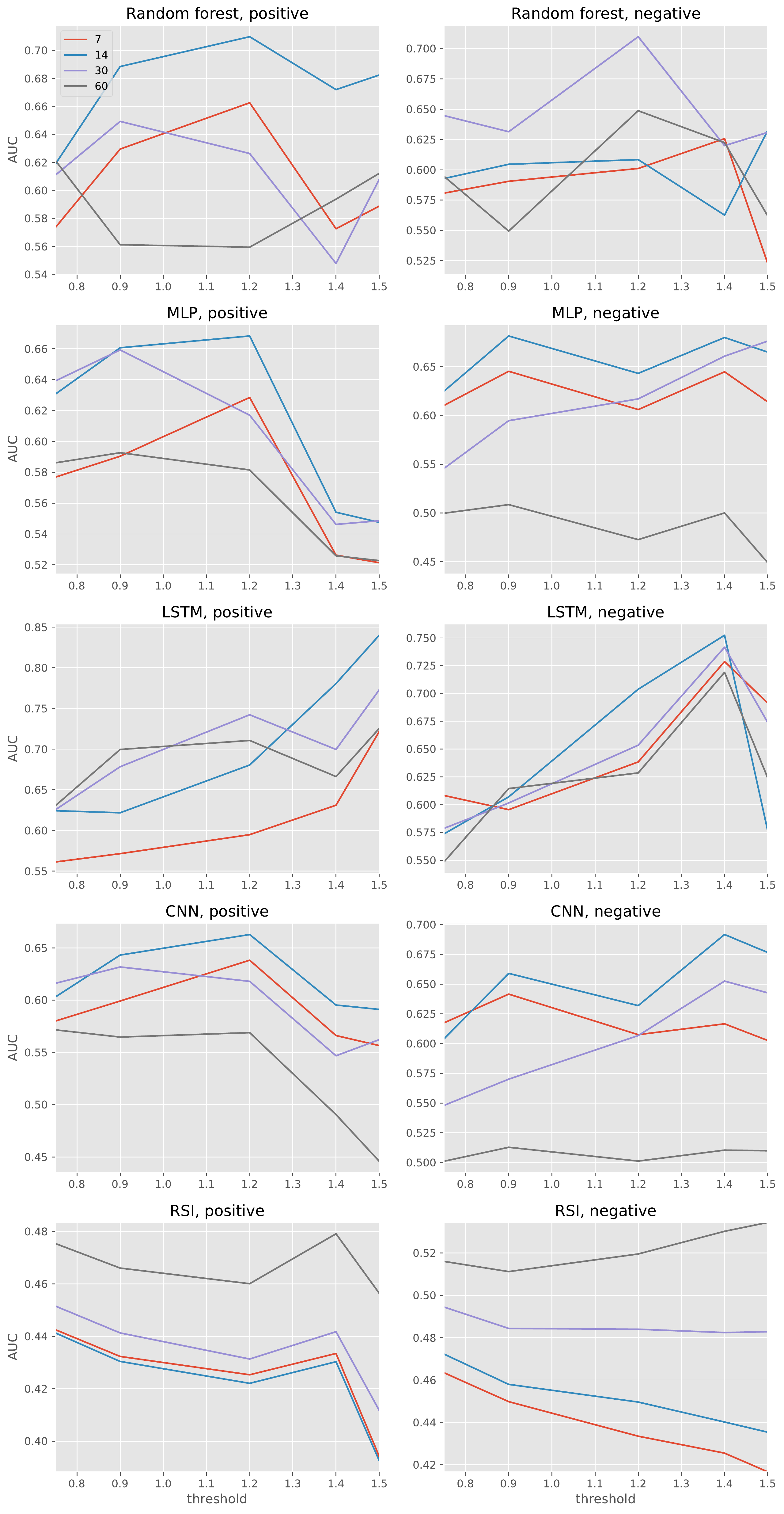}
\caption{Cisco Systems, Inc. stock forecast of significant positive and negative daily returns.}
\label{csco}
\end{figure}

The experiments on Cisco Systems data produce impressive results for the LSTM model.
Our findings are illustrated in Figure \ref{csco}, where the graphs in the first column show performance in forecasting significant positive changes in daily return. Similarly, the graphs in the second column show performance in forecasting significant negative changes. LSTM yields substantially higher AUC values than other models in predicting  positive changes in asset price. It also yields the top results in forecasting negative changes albeit by a smaller margin. We note that LSTM performance improves when forecasting higher changes in price. On the other hand, MLP and CNN models produce different patterns of performance than LSTM. Both MLP and CNN models have poorer performance when predicting higher positive significant changes while a more even performance when predicting negative changes.

\begin{figure}[h!]
\center
\includegraphics[scale=0.42]{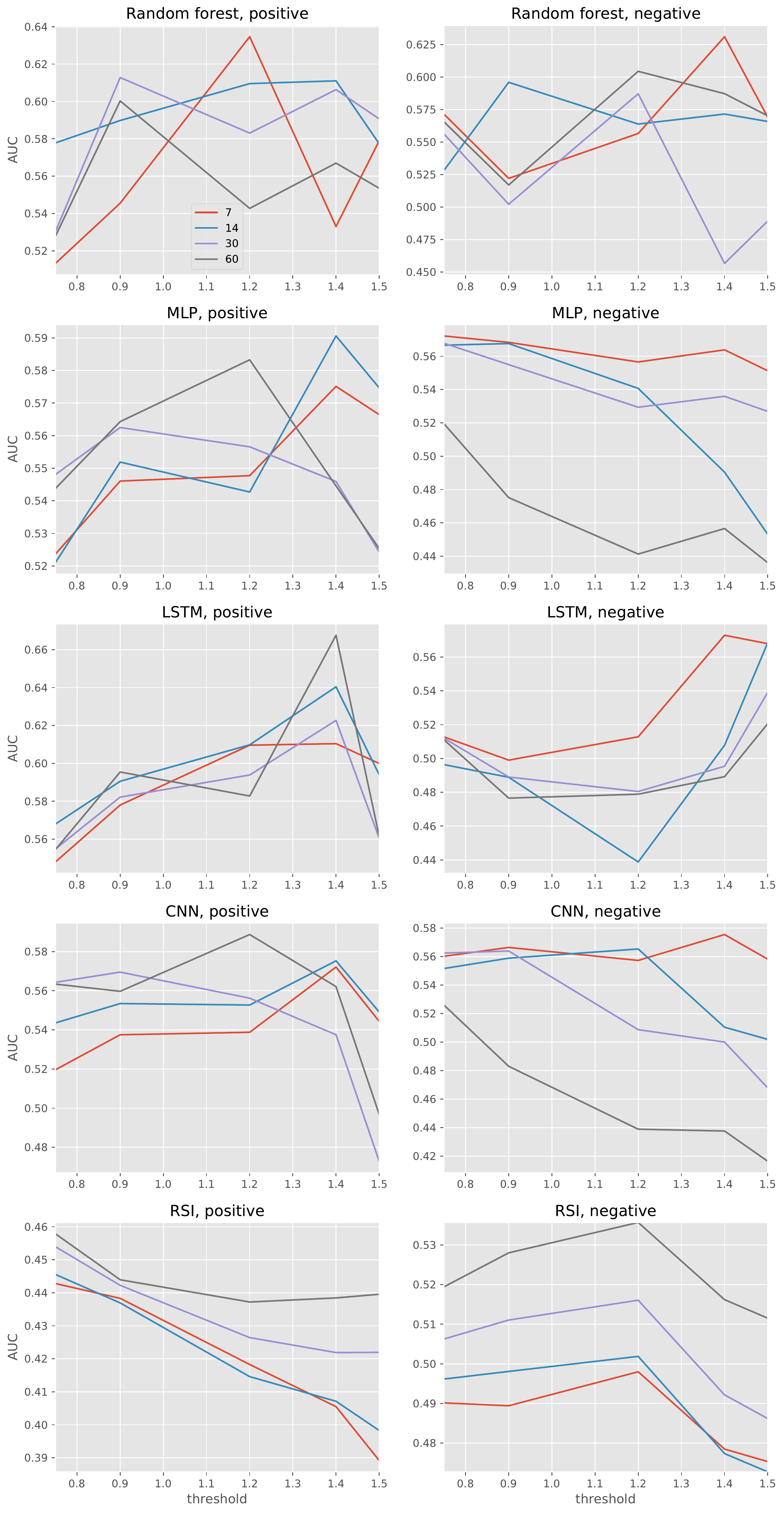}
\caption{The Coca-Cola Company stock forecast of significant positive and negative daily returns.}
\label{ko}
\end{figure}

The experiments on the Coca-Cola Co data yield mixed results as shown in Figure \ref{ko}. 
LSTM produces overall best results in predicting significant positive daily returns. In particular, using a 60 day moving window results in the optimal forecast model. In general, the performance of LSTM improves with increase in significance threshold except for a big dip from 1.4 to 1.5 in the positive case. In addition, using a 7-day window for neural networks yields the best results in forecasting negative changes. On other hand, RF produces overall best result in predicting significant negative returns. However, it is hard to discern any consistent trends in the RF model.

\begin{figure}[h!]
\center
\includegraphics[scale=0.42]{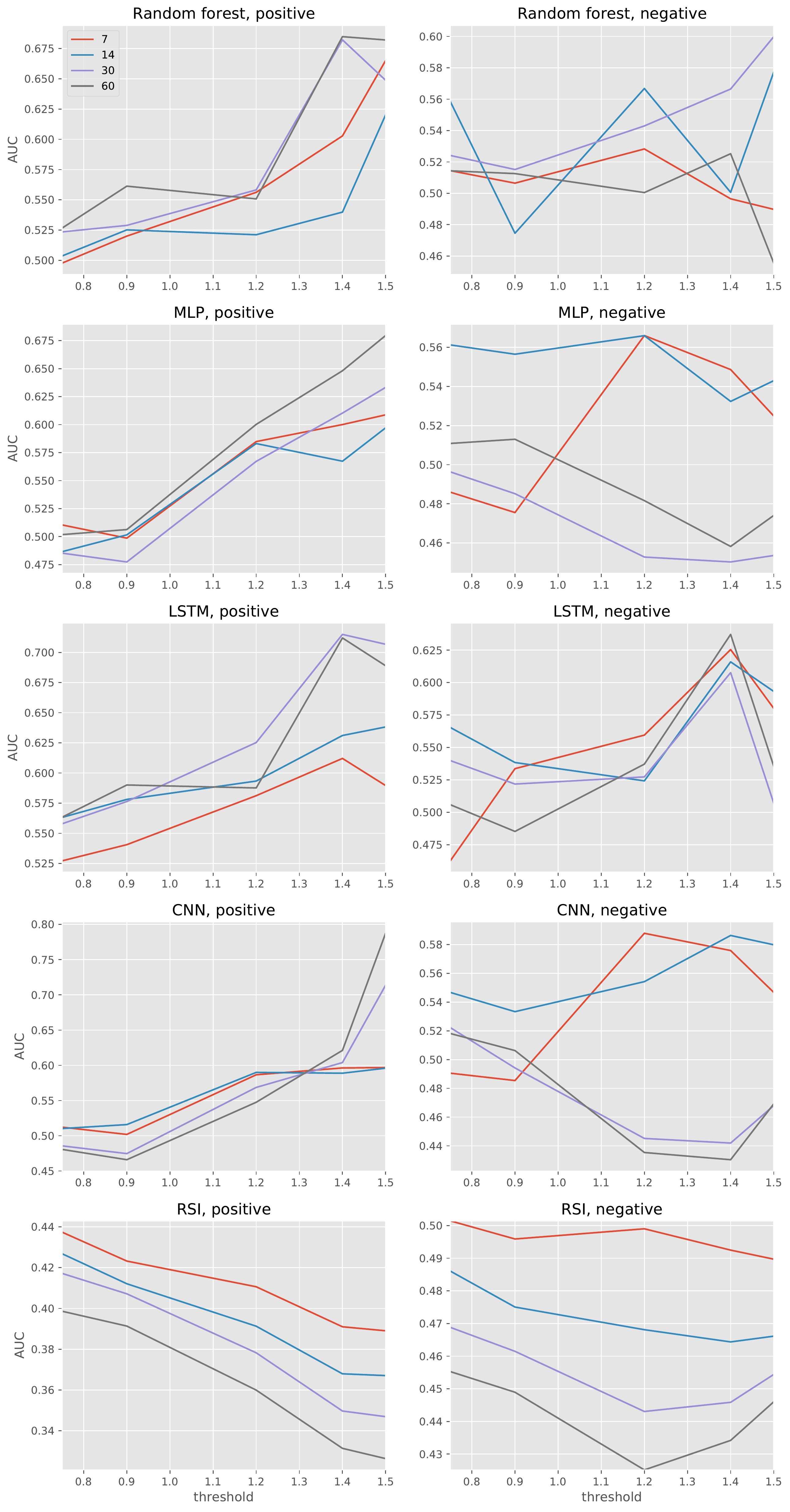}
\caption{Nike, Inc stock forecast of significant positive and negative daily returns.}
\label{ni}
\end{figure}

The experiments on Nike data produce more consistent results as illustrated in Figure \ref{ni}. All four machine learning models show improved performance with increase in the significance threshold when predicting positive daily returns. The performances on the negative return prediction task are less consistent. LSTM has similar graphs in both positive and negative prediction tasks albeit with different AUC values. We also note that in positive return prediction 30 and 60-day window models produce overall better results than shorter term window models. On the other hand, 7 and 14-day models produce better results in negative return prediction.

\begin{figure}[h!]
\center
\includegraphics[scale=0.42]{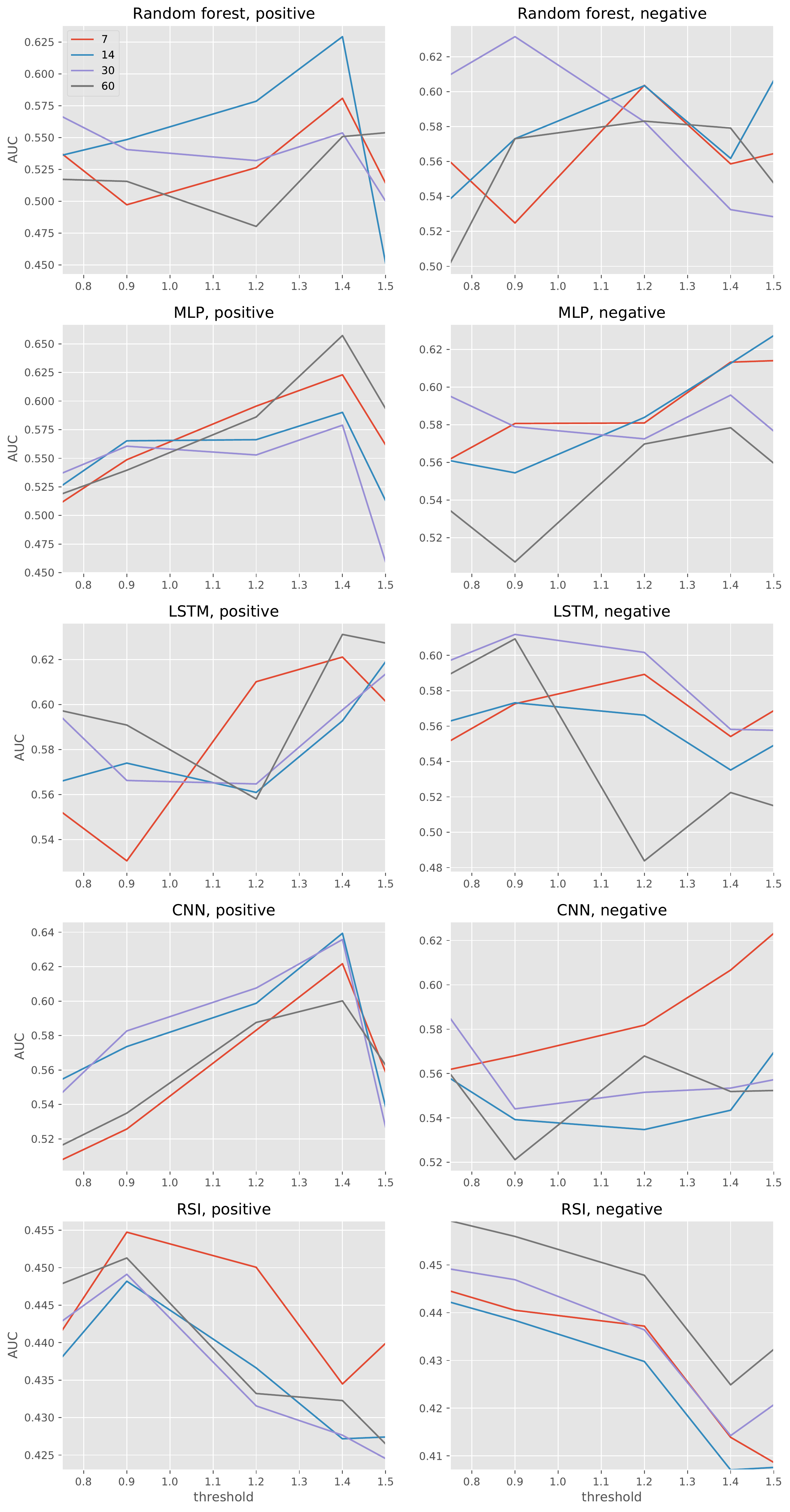}
\caption{The Goldman Sachs Group, Inc. stock forecast of significant positive and negative daily returns.}
\label{gs}
\end{figure}

The experiments on Goldman Sachs data show that in positive return prediction all four machine learning models improve their performance with increase in threshold significance (Figure \ref{gs}). However, the performance generally drops after the threshold of 1.4. This pattern is also observed in other data. We believe that the deterioration in performance can be partially explained by the target class imbalance. Since the number of significant instances decreases with increase of the threshold the target distribution becomes skewed. At the threshold level of 1.5 only about 3\% of the instances are labeled as significant. As a result while the classifier accuracy may improve its AUC deteriorates. It is also plausible that the models simply fail to capture the patterns associated with very big price changes.

%----------------------------------------------------------------------------------------------------------------------------------------------------
%----------------------------------------------------------------------------------------------------------------------------------------------------

\section{Conclusion}

In this paper, we investigate the performance of neural network models in forecasting significant daily returns using previous daily returns. We employed three popular neural net architectures: MLP, CNN, and LSTM. We also used RF and RSI models as benchmarks. The models were tested using 10-year daily price data of four major US public companies. The  companies were chosen to represent a diverse field of industries to avoid correlated results. The data was split temporally for independent training and testing.

The results show that neural network models are capable of forecasting significant changes in asset price with high degree of accuracy as in the case of the LSTM model on the Cisco Systems data. The models' performance generally  improve, up to a certain point, with increase in significance threshold. 
In other words, the models are generally better at predicting more significant changes than less significant ones. We postulate that less significant changes are more random in nature and therefore are harder to model.
Our findings are in line with previous studies  that investigated forecasting the direction of price change which  is equivalent to setting the threshold level to 0.  The studies on predicting the direction of price change obtained AUC results of no more than 0.55 \cite{borovkova, fortuny}.

Although model performance improves with increase in threshold level, in many cases, there is also a significant drop in performance when moving from threshold of 1.4 to 1.5. We attribute this observation partly to class imbalance that occurs when the significance level is very high. Since there are considerably fewer positive observations at very high significance thresholds the response variable distribution becomes skewed which negatively affects the performance of classifiers. Additionally, the price changes at high significance level may be driven by different fundamentals that are not captured by the models.

As mentioned above, the LSTM model is capable of producing superior results in certain scenarios. However, it is a computationally expensive algorithm that requires a long time to train. On the other hand, MLP is relatively fast and is capable of producing competitive results. Therefore, the trade-off between the speed and accuracy must be considered before choosing the best forecast model. We note that RF model also produces robust results. It is computationally very efficient and may serve as a potential alternative to more laborious neural network models.

The models generally improve their performance with increase in significance level. However, there is often a drop in performance at the maximum significance level. This is most likely due to extreme imbalance in class distribution that takes place when there are only a few instance of the minority class. There exist a number of algorithms to balance the data that can be used in the given context \cite{chawla, kamalov}.

The majority of the existing literature in price prediction is focused on the actual price prediction or direction of price change. Our work addresses a previously little explored question of predicting significant changes in asset price. The results show that it is a promising research avenue.
%----------------------------------------------------------------------------------------------------------------------------------------------------
%----------------------------------------------------------------------------------------------------------------------------------------------------
\\
\\
\noindent
\textbf{Conflict of Interest}
\\
\noindent The authors declare that they have no conflict of interest.

%----------------------------------------------------------------------------------------------------------------------------------------------------
%----------------------------------------------------------------------------------------------------------------------------------------------------

\end{document}